\documentclass[aps,pra,twocolumn,superscriptaddress]{revtex4}%
\usepackage{amsmath}
\usepackage{amssymb}
\usepackage{graphicx}
\usepackage{float}
\usepackage{xcolor}
\newcommand{\qed}{\nobreak \ifvmode \relax \else
\ifdim\lastskip<1.5em \hskip-\lastskip \hskip1.5em plus0em
minus0.5em \fi \nobreak \vrule height0.75em width0.5em
depth0.25em\fi} 
\DeclareMathOperator{\tr}{Tr}
\usepackage{braket}

\begin{document}

\title{Exploring the limitations of quantum networking through butterfly-based networks}

\author{Kieran N. Wilkinson}
\affiliation{Computer Science and York Centre for Quantum
Technologies, University of York, York YO10 5GH, United Kingdom}
\author{Thomas P. W. Cope}
\affiliation{Department of Mathematics, University of York, York
YO10 5DD, United Kingdom}
\author{Stefano Pirandola}
\affiliation{Computer Science and York Centre for Quantum
Technologies, University of York, York YO10 5GH, United Kingdom}

\begin{abstract}
We investigate the classical and quantum networking regimes of the
butterfly network and a group of larger networks constructed with
butterfly network blocks. By considering simultaneous multicasts
from a set of senders to a set of receivers, we analyze the
corresponding rates for transmitting classical and quantum
information through the networks. More precisely, we compare
achievable rates (i.e., lower bounds) for classical communication
with upper bounds for quantum communication, quantifying the
performance gap between the rates for networks connected by
identity, depolarizing and erasure channels. For each network
considered, we observe a range over which the classical rate non-trivially
exceeds the quantum capacity. We find that, by adding butterfly
blocks in parallel, the difference between transmitted bits and
qubits can be increased up to one extra bit
per receiver in the case of perfect transmission (identity
channels). Our aim is to provide a quantitative analysis of those
network configurations which are particularly disadvantageous for
quantum networking, when compared to classical communication. By
clarifying the performance of these ``negative cases'', we also
provide some guidance on how quantum networks should be built.
\end{abstract}

\maketitle

\section{\label{sec:intr}Introduction}
\begin{figure}[t]
 \centering
\includegraphics[width=\linewidth]{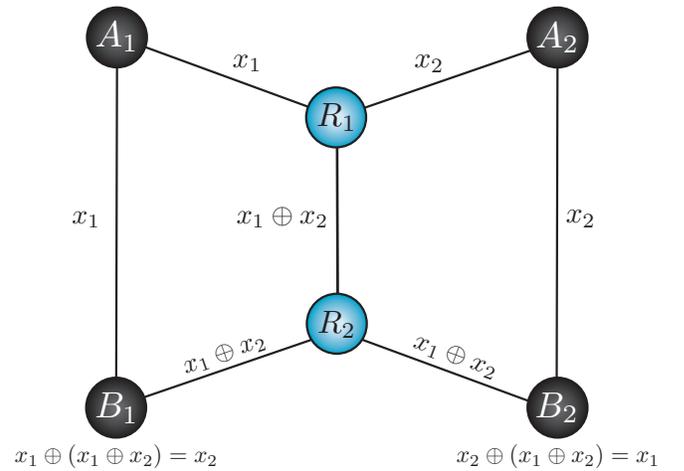}
\caption{A schematic of the butterfly network with two senders,
$A_1$ and $A_2$ and two receivers, $B_1$ and $B_2$. A bottleneck
channel $R_1 \to R_2$ connects each sender to their indirect
receiver. Classical network coding is performed using modulo-2
addition for encoding at $R_1$ and for decoding $B_1$ and $B_2$. Thanks to
network coding, both the single-message multicasts $A_1 \to (B_1,B_2)$ and $A_2 \to (B_1,B_2)$ are simultaneously successful.}
\label{fig:butterflySingle}
\end{figure}

Bottleneck points arise frequently in network topology. One of the
simplest examples of the bottleneck problem is in the butterfly
network as shown in Fig.~\ref{fig:butterflySingle}. Consider the
two-pair communication problem in which the pair of senders $A_1$
and $A_2$ perform single-message multicasts to the pair of receivers $B_1$ and
$B_2$ under the flooding condition that data can only be sent
through unused connections. We encounter a bottleneck at node
$R_1$ as data is waiting to be sent from both senders through the
channel $(R_1, R_2)$. In 2002, a solution was proposed by Ahlswede
et al. in which the bottleneck problem could be bypassed using
network coding \cite{Ahlswede2000-ay}.  As outlined in
Fig.~\ref{fig:butterflySingle},  network coding in the butterfly
network is performed by encoding data using an XOR operation at
$R_1$ and decoding with a second XOR operation at the receivers,
after transmission through the bottleneck channel. The result is
that each sender successful multicasts their data to both
receivers with only a single use of each channel.

In general (and in our work), one may consider the general case
where the multicasts can be partially achieved. In fact, in a
noisy version of the network, we may associate an average rate to
each sender which accounts for the fact that sometimes only a
subset of the receivers is reached. In this context we are
interested in the average number of bits per receiver that
can be transmitted from the senders in each multicast use of the
network.

This problem can be analyzed in the setting of quantum information
theory~\cite{Watrous,Bengtsson,Nielsen2002-sc,Preskill1998-ts,HolevoBOOK,SamRMP,RevPirs,hybrid},
e.g., connected to the general aim of enabling a wide-spread
implementation of a quantum
internet~\cite{rep1,rep2,rep3,Pirandola2016-ke, Kimble2008-yb,
Meterbook, telereview} or to build large communication networks
for quantum key distribution~\cite{revQKD}. There are several
limitations of quantum mechanics that become important in network
implementations, most notable the inability to perfectly clone
quantum information~\cite{noclone1, Wootters1982-ys}. It is this
property that prevents network coding in the quantum form of the
butterfly network. Ref.~\cite{Hayashi2007-fk} confirmed that there
is no quantum process which achieves network coding with unit
fidelity while demonstrating that approximate quantum network
coding can be achieved using universal cloning ~\cite{Buzek1996-is},
obtaining a fidelity greater than 1/2 but no more than 0.983.
Later, Ref.~\cite{Jain2011-ph} provided an information-theoretic
proof that quantum network coding does not provide a larger
information flow than routing in the butterfly network.


Perfect quantum network coding has been shown possible in the setting of quantum state transfer in perfect channels. However, achieving this requires the use of additional resources, for example, Ref.~\cite{Hayashi2007-tl} showed that perfect network coding can be achieved if the senders share an entangled state before transfer begins. Several other protocols make use of prior entanglement to achieve the same goal~\cite{Satoh2012-dz, Nishimura2013-gm, Li2019-bz}. Furthermore, a method of quantum computing which utilizes the prior entanglement in the butterfly network to perform unitary operations without the need for further entanglement resources has been developed~\cite{Soeda2011-gs}. Alternatively, Ref.~\cite{Kobayashi2009-fn} has shown that transfer of quantum states by quantum network coding is possible in the absence of prior entanglement by enabling free classical communication between nodes and several other investigations have considered a free classical communication regime~\cite{Kobayashi2011-va, Leung2010-wc,Nishimura2013-gm, Luo2014-eg}.



The majority of the literature concerning quantum network coding in the butterfly network has focused on an information processing setting in which quantum states are simply transferred or reconstructed. Here, we consider the problem in a quantum communication setting, in which the quantum systems are physically sent through quantum channels affected by environmental noise. In fact, we investigate the classical and quantum communication rates of the butterfly network based on identity, depolarizing and erasure channels. We build on previous results from Refs.~\cite{Pirandola2017-xo,QSTpirs,Pirandola2016-xl,PirNET,PirMultiNET} to upper bound the highest quantum communication rates achievable in (single-message) multicasts from senders to receivers, assisted by adaptive local operations (LO) and two-way classical communication (CC) involving all the nodes of the network. We then compare these bounds with the corresponding rates that can be achieved for (single-message) multiple multicasts of classical information from senders to receivers, establishing parameter regimes where classical outperforms quantum communication. This analysis is then extended to larger networks built on butterfly network blocks, for which this gap can be amplified. In this way, our study clarifies the non-trivial limitations that certain network architectures have for transmitting quantum information.



\section{\label{sec:res}Preliminary notions}

Before presenting our main results, let us introduce some necessary notions by
reviewing some of the theory developed in
Refs.~\cite{Pirandola2017-xo,QSTpirs,Pirandola2016-xl,PirNET,PirMultiNET}. These works
consider protocols for quantum transmissions, entanglement distribution
and secret key generation which are based on the most general LOs
assisted by two-way CCs among all the parties involved, that we may briefly
call adaptive LOCCs. In particular, Ref.~\cite{Pirandola2017-xo} considered
point-to-point protocols over a quantum channel, while Ref.~\cite{PirNET}
extended the study to protocols over repeater chains and, more generally,
quantum networks. Finally, Ref.~\cite{PirMultiNET} further extended the study
to quantum communication networks with multiple senders and receivers.

Using this theory, it is possible to
write an upper bound for the total number of qubits that an
ensemble of senders can transmit to an ensemble of receivers in terms of the relative entropy of entanglement (REE), defined as
\begin{equation}
	E_R(\sigma) = \inf_{\gamma\in \text{sep}} S(\sigma||\gamma)
\end{equation}
where the infimum is over all the set of separable states (sep) and $S$ is the quantum relative entropy~\cite{vedral}
\begin{equation}
	S(\sigma||\gamma) := \tr\left[\sigma\left(\log_2\sigma - \log_2\gamma\right)\right].
\end{equation}

In this work we consider the identity, depolarizing and erasure channels, all of which belong to the group of teleportation covariant channels. A channel $\mathcal{E}$ belongs to this group if, for any teleportation unitary $U$, we may write
\begin{equation}
	\mathcal{E}(U\rho U^\dagger) = V\mathcal{E}(\rho)V^\dagger
\end{equation}
for another (generally different) unitary $V$. For a teleportation covariant channel $\mathcal{E}$, we may write a single-letter REE bound for its two-way quantum capacity $Q_2$, i.e.~\cite{Pirandola2017-xo}
\begin{equation}
	Q_2(\mathcal{E}) \leq E_R(\mathcal{E}) =  E_R(\sigma_\mathcal{E})
\end{equation}
where $E_R(\mathcal{E}) = \max_\sigma E_R\left[(\mathcal{I}\otimes\mathcal{E})(\sigma)\right]$,
and $\sigma_\mathcal{E} = (\mathcal{I}\otimes\mathcal{E})(\Phi )$ is the Choi matrix of the channel, with $\Phi$ being the maximally entangled state. (The proof that the Choi matrix maximizes the REE of a teleportation covariant channel is given in Ref.~\cite{Pirandola2017-xo})

In order to generalize these results to teleportation-covariant networks we first introduce some notions from graph theory~\cite{Pirandola2016-xl,PirNET,PirMultiNET}. Let us describe an arbitrary quantum network $\mathcal{N}$ as an
undirected graph with nodes or points $P$ and edges $E$. Two
points, $x$ and $y$, are connected by an edge $(x,y) \in E$ if and
only if there is a corresponding quantum channel
$\mathcal{E}_{xy}$ between the two. Each point $p$ has a local
register of quantum systems over which LOs are performed and
optimized on the basis of two-way CCs with the other nodes. Given
a set of senders or Alices $\{A_i\}$ and a set of receivers or
Bobs $\{B_j\}$, we define a cut $C$ as a bipartition
$(\mathbf{A},\mathbf{B})$ of the points $P$ such that $\{A_i\}
\subset \mathbf{A}$ and $\{B_j\} \subset \mathbf{B}$. This is also
denoted as $C:\{A_i\}|\{B_j\}$. Then, a cut-set $\tilde{C}$
corresponds to the set of edges $(x,y)$ which are disconnected by
the cut $C$, i.e., such that $x \in \mathbf{A}$ and $y \in
\mathbf{B}$.

Let us first consider the unicast configuration where a single sender (Alice, $A$) communicates with a single receiver (Bob, $B$) over a quantum network. If the communication is done via a single route, we can obtain an upper bound on the quantum capacity using the single-edge flow of REE through a network. For a cut $C:A|B$ with cutset $\tilde{C}$ and Choi matrix resource state $\sigma_{\mathcal{E}_{xy}}$ for edge $\mathcal{E}_{xy}$, this flow is defined by
\begin{equation}
	E_R(C) := \max_{(x,y)\in \tilde{C}}E_R(\sigma_{\mathcal{E}_{xy}}).
\end{equation}
It can then be shown that the single-path (two-way assisted) quantum capacity of the network is upper-bounded as~\cite{PirNET}
\begin{equation}
	Q_2({\mathcal{N}})\leq\min_{C:A|B} E_R(C).
\end{equation}

Alternatively, the parties may choose to use a flooding protocol in which all of the edges of the network are used exactly once by simultaneous routing from Alice to Bob. In this case quantity of interest is the multi-edge flow of REE through cut $C:A|B$ defined by
\begin{equation}
	E_R^m(C) = \sum_{(x,y)\in \tilde{C}}E_R(\sigma_{\mathcal{E}_{xy}})
\end{equation}
which leads to the following upper bound on the multipath (two-way assisted) quantum capacity~\cite{PirNET}
\begin{equation}
	Q_2(\mathcal{N}) \leq \min_{C:A|B} E_R^m(C).
	\label{multicastbound}
\end{equation}

The upper bound in Eq.~(\ref{multicastbound}) provides the maximum number of qubits that Alice can send to Bob per `parallel' use of the quantum network, where all its edges are simultaneously exploited. Here we note that this upper bound can be modified to bound the total number of qubits that an ensemble of Alices $\{A_i\}$ can send to an ensemble of Bobs $\{B_j\}$. In fact,
it is sufficient to consider the two ensembles as super-users and repeat the reasoning, but with the difference that the previous cuts $C:A|B$ must now be replaced by cuts splitting the super-users, i.e., the two ensembles, $C:\{A_i\}|\{B_j\}$. This is
an upper bound because the super-users may in principle apply non-local quantum operations among their nodes and, therefore, better optimize the transmission rate with respect to the case of ensembles of separate users.
As a result, the optimal rate at which qubits can be transmitted from the senders to the receivers is bounded by
\begin{equation}
    \mathcal{B}(\mathcal{N}) := \min_{C:\{A_i\}|\{B_j\}}  \sum_{(x,y)\in \tilde{C}}E_R(\sigma_{\mathcal{E}_{xy}}).\label{eq18bb}
\end{equation}

It is also important to note that this is a general bound for multiple multicasts which applies to both the case of single- and multi-message multicasts from senders to receivers. In fact,
since we bound the total number of physical qubits that super Alice transmits to super Bob, it does not matter if these qubits are independent (i.e., in a tensor product of different states) or dependent (e.g., in a global GHZ state~\cite{GHZ})
when we unravel super Bob back into an ensemble of Bobs. In this work, we are particularly interested in single-message multiple multicasts, where in each use of the network each Alice
aims to send a GHZ-like multipartite logical qubit to the destination set. This is a logical qubit $\alpha |\bar{0} \rangle + \beta |\bar{1} \rangle$  encoded into physical qubits as many as the receivers, i.e., $|\bar{0} \rangle = |0...0 \rangle$ and $|\bar{1} \rangle = |1...1 \rangle$.
In this context, the total number of logical qubits that are correctly received by the destination set is equal to the total number of physical qubits correctly received by each
individual receiver, which means that we need to divide the bound in Eq.~(\ref{eq18bb}) by the number of receivers $r$. Therefore our figure of merit
is the total number of qubits per use and receiver, which is less than or equal to the quantum bound
\begin{equation}
    R_Q(\mathcal{N}) := r^{-1} \min_{C:\{A_i\}|\{B_j\}}  \sum_{(x,y)\in \tilde{C}}E_R(\sigma_{\mathcal{E}_{xy}}).\label{eq18b}
\end{equation}

\section{\label{ssec:singleBut}Rates of a single butterfly block}

By using the bound in Eq. (\ref{eq18b}), we can see immediately
that we cannot obtain more than $1.5$ qubits per use and receiver
in a butterfly network built with identity channels (for which $E_R(\sigma_{\mathcal{E}_{xy}})$ is equal to one qubit).
This comes from the fact that the minimum cut is the middle one disconnecting the edges $(A_1,B_1)$,
$(R_1,R_2)$, and $(A_2,B_2)$. This gives $3$ physical qubits to be divided by $r = 2$.
By contrast, we know from network
coding theory that in the classical case we can obtain two
classical bits per use and receiver; hence we have a difference of $0.5$ bit
per use and receiver between the quantum and classical networks in this case.

We can now consider more complicated and realistic cases in which
the butterfly network is constructed with noisy channels. Let us
start with the depolarizing channel whose action in $d$ dimensions
can be expressed as
\begin{equation}
     \mathcal{P}_\text{depol}(\rho) = (1-p)\rho + \frac{p}{d} \mathcal{I},
\end{equation}
 where $\rho$ is an arbitrary density matrix and $\mathcal{I}$ is the identity matrix. The action of the channel is to transmit the initial state with probability $1-p$ or a maximally mixed state with probability $p$ (geometrically, the action of the depolarizing channel can be thought of as shrinking the Bloch sphere \cite{Nielsen2002-sc}). At the time of writing the exact two-way quantum capacity of the depolarizing channel is unknown, however Ref.~\cite{Pirandola2017-xo} obtained an upper bound of
\begin{equation}
    Q_2(p) \leq E_R(\sigma_{\mathcal{P}_\text{depol}}) = 1-H_2\bigg(1-\frac{3p}{4}\bigg)
\end{equation}
for $p\leq 2/3$ with $Q_2=0$ otherwise, where $H_2(p) = -p\log_2p
- (1-p)\log_2(1-p)$ is the binary Shannon entropy. Thus, for a
butterfly network $\mathcal{B}_{\mathrm{dep}}$ connected by
depolarizing channels with equal probability $p$, the total number
of qubits per use and receiver is bounded
by
\begin{equation}
 R_Q(\mathcal{B}_{\mathrm{dep}}) = \frac{3}{2}\left[ 1-H_2\bigg(1-\frac{3p}{4}\bigg)\right].
\end{equation}

The classical capacities of the depolarizing channel have been
extensively studied~\cite{Mulherkar2016-pc, Wouters2009-hd,
King2003-eg}. The unassisted classical capacity is given by
\begin{equation}
 C(p) = 1 - H_2\bigg(1-\frac{p}{2}\bigg).
\end{equation}
This can be better understood by propagating an encoded classical
bit through the channel. For  the input $\ket{0}\bra{0}$, we get
\begin{align}
    \mathcal{P}(\ket{0}\bra{0}) &= (1-p)\ket{0}\bra{0} + \frac{p}{2} (\ket{0}\bra{0} + \ket{1}\bra{1}) \nonumber \\
    &= \bigg(1-\frac{p}{2} \bigg) \ket{0}\bra{0} + \frac{p}{2} \ket{1}\bra{1} ,
\end{align}
similarly for $\ket{1}\bra{1}$,
\begin{equation}
    \mathcal{P}(\ket{1}\bra{1}) = \bigg(1-\frac{p}{2} \bigg) \ket{1}\bra{1} + \frac{p}{2} \ket{0}\bra{0}.
\end{equation}
What we have is the equivalent of a classical binary symmetric
channel (BSC) with bit flip probability $p/2$~\cite{Cover}.

\begin{figure}[t!]
 \centering
\includegraphics[width=\linewidth]{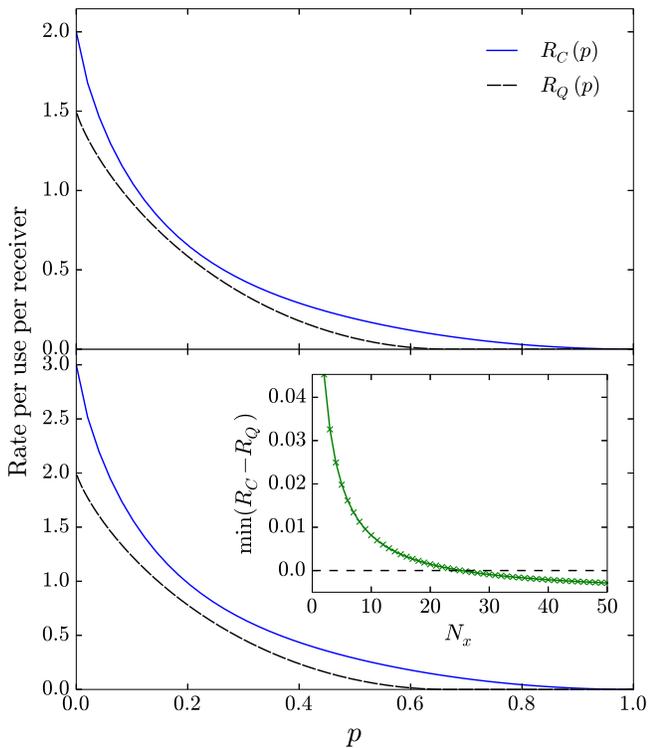}
\caption{Rates (bits/qubits) per use and receiver for the depolarizing case considering a single
butterfly block (upper panel) and the limit of $N_x\to\infty$
blocks in parallel (lower panel). We plot the achievable classical rate $R_C$ (solid blue line) and the quantum bound $R_Q$ (dashed black line).
The inset shows the minimum difference between the classical rate and the quantum bound as a function of $N_x$ (where the minimization is taken over the probabilities).} \label{fig:depolCaps}
\end{figure}

\begin{figure}[t!]
 \centering
\includegraphics[width=\linewidth]{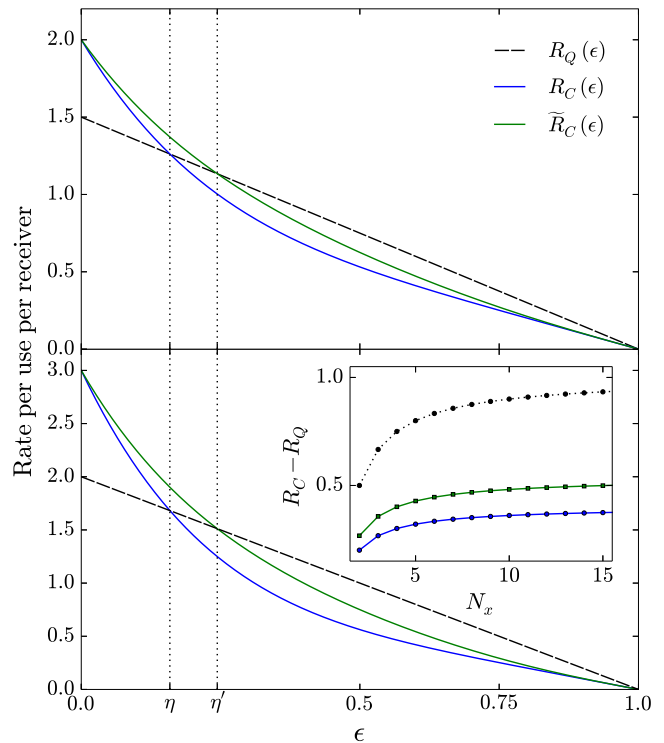}
\caption{Rates (bits/qubits) per use and receiver for the erasure case, considering a single
butterfly block (upper panel) and the limit of $N_x\to\infty$
blocks in parallel (lower panel). We plot the quantum bound $R_Q$ (dashed black line), the achievable classical rate $R_C$ (solid blue line) and the inter-node-assisted achievable
classical rate $\tilde{R}_C$ (solid green line). Note that the maximum gap is achieved for $\epsilon = 0$ where the channels
become identity channels. At that point, for a single butterfly block, we find the expected result of 2 bits versus 
1.5 qubits per use per receiver. In the figure, $\eta$ and $\eta'$ are the critical
points at which $R_C$ and $\tilde{R}_C$, respectively cross the
quantum bound $R_Q$. Inset: Difference between $\tilde{R}_C$ and
$R_Q$ as a function of $N_x$ for values of
erasure probability equal to 0 (black line) and $\eta'/2$ (green
line), and the difference between $R_C$ and $R_Q$ as a function of $N_x$ for erasure probability equal to $\eta/2$ (blue line).} \label{fig:erasureCaps}
\end{figure}

To compute an achievable rate for the classical single-message multiple multicast
over a depolarizing/BSC butterfly network, we deconstruct the network in the two channels
$A_1,A_2\to B_1$ and $A_1,A_2\to B_2$. Calculating the total rate
of the combined channels gives an achievable rate for the network.
The general procedure for this process is to create the transition
probability matrix using the logic of the butterfly network,
followed by an optimization over a distribution on the input
symbol. The upper panel of Fig.~\ref{fig:depolCaps} shows both the
quantum bound $R_Q$ (qubits per per use and receiver) and the
achievable rate $R_C$ for sending classical information (bits per use and receiver).
The quantum bound is exceeded by the classical rate over the entire range of $p$ with the maximum
difference being 0.5 bits per receiver (which corresponds to the ideal case of identity channels
discussed above).


Let us now move on to erasure channels. From classical information
theory we know that the capacity of the binary erasure channel
with erasure probability $\epsilon$ is given by
$C(\epsilon)=1-\epsilon$. This formula has also been shown to be
equal to the classical capacity of the quantum erasure
channel~\cite{Bennett1997-pb}. The same work found the quantum
capacity to be $Q(\epsilon)=\max\{0,1-2\epsilon\}$ and also
$C(\epsilon)= Q_2(\epsilon) = 1-\epsilon$. 
The erasure channel is unique in that
the number of correctly transmitted bits is known with certainty, so that the capacity is equivalent to the average number of
transmitted bits. For any network it is straightforward to calculate the
achievable classical rate. For a single butterfly network block
$\mathcal{B_{\mathrm{era}}}$ connected by erasure channels with
the same probability, we obtain the classical rate (per use and receiver)
\begin{equation}
    R_C(\mathcal{B}_\mathrm{era})= (1-\epsilon) + (1-\epsilon)^5,
    \label{CSingle}
\end{equation}
where the first term arises from the contribution of the side
channels, and the second comes from network coding at the
bottleneck node.

Allowing side one-way CC between nodes in the network allows for
the optimization of the transmission routes, increasing the rate.
For example, if we detect a failure in a channel connected to the
bottleneck node, $(A_i, R_1)$, we send any data received at $R_1$
directly to $R_2$ and subsequently to both receivers. We then have
additional communication paths from $A_j$ to $B_i$ and $B_j$. The
inter-node-assisted rate (per use and receiver) is given by
\begin{equation}
    \tilde{R}_C(\mathcal{B}_\mathrm{era}) = (1-\epsilon) + (1-\epsilon)^5 + \epsilon(1+\epsilon)(1-\epsilon)^3.
\end{equation}

The upper panel of Fig.~\ref{fig:erasureCaps} shows each of the
rates for a single butterfly block. For both $R_C$
and $\tilde{R}_C$, we observe a region where the quantum bound
$R_Q$ is exceeded. We label the crossing points $\eta =0.159$ and
$\eta'=0.244$ for $R_C$ and $\tilde{R}_C$, respectively. The
advantage of inter-node classical communications is significant,
and extends the performance difference between the classical and
quantum butterfly network in this configuration.

\section{Building networks with butterfly blocks}
\begin{figure}[ht!]
  \centering
\resizebox{0.48\textwidth}{!}{
\includegraphics{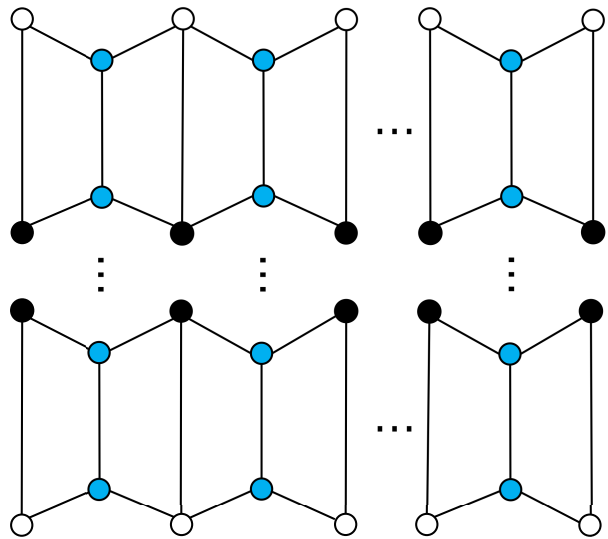}
} \caption{Diagram of the construction of larger networks from
butterfly network blocks in parallel (horizontal) and series
(vertical).} \label{fig:xxy}
\end{figure}

We will now expand our analysis to larger networks constructed
with butterfly network blocks as shown in Fig.~\ref{fig:xxy}. We
consider adding blocks in parallel in Sec.~\ref{sec:parallel}, in
series in Sec.~\ref{sec:series} and in both series and parallel in
Sec.~\ref{sec:both}.

\subsection{\label{sec:parallel}Butterfly blocks connected in parallel}
Firstly, we consider the case in which we have a single row of $N_x$
connected butterfly blocks, such that we have a network
$\mathcal{N}_\mathrm{par}$ with $r=N_x+1$ senders/receivers. We
extend the previous reasoning to evaluate the maximum number
of multipartite logical qubits that can be flooded/transmitted from senders to
receivers per use of the network. This is equal to the total number of
physical qubits transmitted per use and receiver, for which we can write corresponding
upper bounds. In particular, for the depolarizing case, we have the quantum bound
\begin{equation}
    R_Q = \frac{2r-1}{r}\bigg[1-H_2\bigg(\frac{3p}{4}\bigg)\bigg],
\end{equation}
while for the erasure case we write
\begin{equation}
    R_Q = \frac{2r-1}{r}(1-\epsilon).
\end{equation}

The achievable rate of the depolarizing network can be found
by expanding the methods used for the single butterfly block case.
The network can be deconstructed into two channels of the form
$A_i,A_{i+1}\to B_i$ at the ends of the network, and $\left(N_x-1\right)$
channels of the form $A_i,A_{i+1},A_{i+2}\to B_{i+1}$. We find the
overall rate numerically from the combination of the capacities of
all channels, from which we can compute the rate per user and receiver $R_C$.
For the erasure case we can directly write the following rate per user and receiver
\begin{equation}
    R_C = (1-\epsilon) + \frac{2(r-1)}{r}(1-\epsilon)^5.
    \label{eqn:generalErasRow}
\end{equation}
Here the first term on the right hand side is due to the fact that
all receivers may receive a single bit from their directly
connected sender, while the second term accounts for the fact that
all receivers (except the extremal ones $B_1$ and $B_r$) may
receive two bits from successful network coding on the adjacent
intermediate nodes. In the case of inter-node CC, the rate can be
generalized by recognizing that we have Eq.~(\ref{eqn:generalErasRow}) plus four backup routes per butterfly
block, giving an overall rate of
\begin{equation}
    \tilde{R}_C = (1-\epsilon) + \frac{2(r-1)}{r}\big[(1-\epsilon)^5 + \epsilon(1+\epsilon)(1-\epsilon)^3\big].
\end{equation}

We see immediately for the erasure network that the difference
between the average number of bits/qubits grows monotonically as we
increase the number of butterfly blocks. Taking the limit of large
$r$ for the rates of the erasure case we obtain
\begin{align}
    &\lim_{r\to \infty} R_Q = 2(1-\epsilon)\\
    &\lim_{r\to \infty} R_C = (1-\epsilon) + 2(1-\epsilon)^5\\
    &\lim_{r\to\infty} \tilde{R}_C = (1-\epsilon) + 2(1-\epsilon)^5 + 2\epsilon(1+\epsilon)(1-\epsilon)^3.
\end{align}

The lower panel of Fig.~\ref{fig:depolCaps} shows the rates of the
depolarizing case in the limit of large $r$ for the entire range
of probabilities. The asymptotic rates are
approximately identical at 0.2 but the classical case outperforms
the quantum bound everywhere else in the range. The lower panel of
Fig.~\ref{fig:erasureCaps} shows the asymptotic rates for the erasure case.
We find that $\eta$ and $\eta'$ are
equivalent to the single block case for any non-zero $N_x$. At small values of the erasure probability
$\epsilon$, the gap between the rates tends to
one bit per use per receiver.


\subsection{\label{sec:series}Butterfly blocks connected in series}
We now consider the rates of a network
$\mathcal{N}_{\mathrm{ser}}$ consisting of $N_y$ butterfly network
blocks connected in series i.e. in a ladder formation. The number of receivers is the same ($r=2$) but the number of intermediate nodes and channels has increased.
The addition of extra blocks has the effect of reducing the rates, as
it becomes harder to reach a receiver without the incurrence of
errors.

\begin{figure}[h!]
 \centering
\resizebox{0.48\textwidth}{!}{
\includegraphics{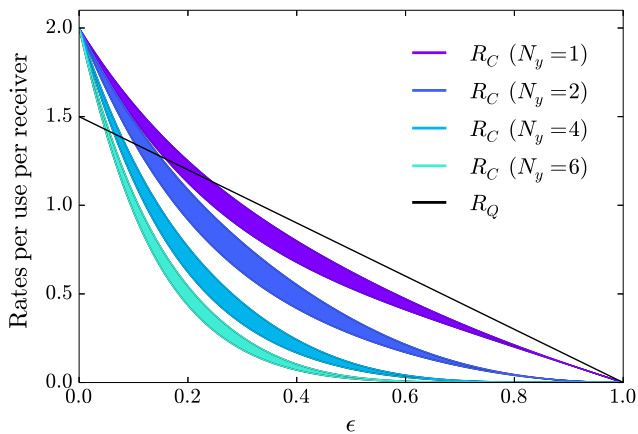}
} \caption{Classical rate (lower lines) and inter-node CC assisted
classical rate (upper lines) per receiver for a $1\times N_y$
butterfly block network constructed with erasure channels. Comparison with the quantum bound (solid black line).}
\label{fig:erasureSeries}
\end{figure}

For the depolarizing case, adding blocks in a ladder structure is
equivalent to adding only extra side channels above a single
block. However, in the erasure case, extra bottlenecks can be used
effectively, even if there are no additional communications. If we
allow the nodes to duplicate data, we can use the bottleneck
channels as effective backup channels in case of errors and
perform network coding in the final bottleneck before the
receivers.

 Let us firstly consider only two blocks in series. We have the option to ignore the bottleneck structure in the first block entirely and send only via the side quantum channels. In this case the classical rate (per use and receiver) is given by
\begin{equation}
    R_C = (1-\epsilon)^2 + (1-\epsilon)^7
\end{equation}
which is however not optimal. A more effective strategy is to use
the bottleneck as an effective backup channel, by sending a bit
from $A_1$ to the intermediate node on $A_1$'s side of the network, which we briefly label $I_1$, via the channel $R_1\to R_2$ as well as through the direct channel $A_1\to I_1$. Now $I_1$ has a greater probability of receiving the correct bit, and because there are no additional operations, no communication between nodes is required. We can calculate the probability that a correct bit is received by computing
\begin{equation}
   \lambda = 1-P(\text{fail}) = 1-\epsilon[1-(1-\epsilon)^3]
\end{equation}
and the classical rate (per use and receiver) is therefore given by
\begin{align}
    R_C= \frac{(1-\epsilon)\lambda +(1-\epsilon)^2}{2} + (1-\epsilon)^6\lambda.
\end{align}
There is no further way we can improve the rate without the
receivers losing certainty of what has been sent.

This strategy
can be extended to any number of blocks in series, where a backup
channel can be applied once per block. Sender/intermediate nodes
on either side of the network can use the bottleneck route,
however, for more than two blocks the rate is
maximized when the routes are always used by nodes on the same
side of the network. The previous classical rate can be generalized as
\begin{equation}
	R_C=\frac{(1-\epsilon)\lambda^{N_y-1}+ (1-\epsilon)^{N_y}}{2} + (1-\epsilon)^5(1-\epsilon)^{N_y-1}\lambda^{N_y-1}.
\end{equation}

Alternatively, if we allow inter-node communication, the classical rate of a $1\times N_y$ erasure network is obtained by considering all of the possible paths from sender to receiver, while prioritizing the backup route in upper blocks, and accounting for possible channel failures. The classical rates for the $1\times N_y$ network are shown in Fig.~\ref{fig:erasureSeries} and compared with the quantum bound $R_Q$ which does not depend on $N_y$. Clearly the value of the crossing point $\eta'$ decreases rapidly as $N_y$ increases, but there is still a significant gap between the upper bound on the quantum rate and the achievable classical rate.

\subsection{\label{sec:both}Butterfly blocks connected in series and parallel}

Finally, we come to the most complex case in which we consider a
general $N_x\times N_y$ grid of butterfly network blocks. This means that
we have $r=N_x+1$ receivers. Again,
we calculate the classical rates, accounting for how the
additional bottlenecks may be exploited. By allowing each sender
(excluding the one at the right edge of the network) to use the backup route
to the its right in $(N_y-1)$ upper blocks, we obtain the following unassisted classical rate (per use and receiver)
for a general grid
\begin{align}
	&R_C = \frac{1}{N_x+1} \{N_x(1-\epsilon)\lambda^{N_y-1} + (1-\epsilon)^{N_y}\nonumber \\
	&+ 2(N_x-1)(1-\epsilon)^5\lambda^{2(N_x-1)}\nonumber \\
    &+ 2(1-\epsilon)^5(1-\epsilon)^{N_y-1}\lambda^{N_y-1}\}.
\end{align}

\begin{figure}[ht!]
 \centering
\resizebox{0.48\textwidth}{!}{
\includegraphics{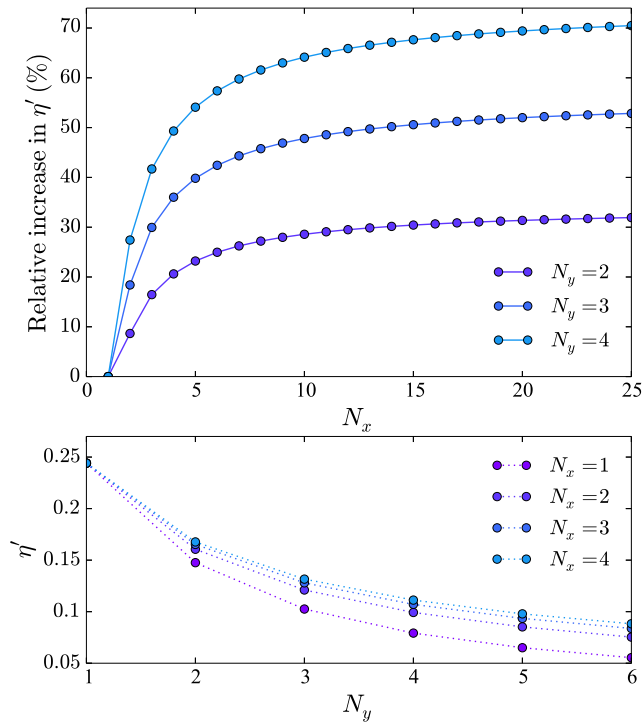}
} \caption{Upper panel: relative increase in the critical point
$\eta'$ as compared to the $N_x=1$ case for various values of
$N_y$. Lower panel: variation of $\eta'$ with $N_y$ for various
values of $N_x$.} \label{fig:XxYAnalysis}
\end{figure}

For the inter-node assisted rate, we repeat the strategy of the series-only case and obtain values of $\eta'$ for different configurations. The top panel of Fig.~\ref{fig:XxYAnalysis} shows the relative increase in the critical point $\eta'$ with respect to the
series-only case. The increase is significant and increases with
the number of blocks we have in series. The lower panel shows
$\eta'$ as a function of $N_y$. The point $\eta'$ decreases
rapidly as we increase the number of blocks between sender and
receiver, however the results show that we always have a finite
range over which the classical rate exceeds the quantum bound.
These results demonstrate that by adding more blocks in parallel
we can increase $\eta'$ up to a convergence point, increasing by
more than 60\% in some cases.

\section{Conclusions}

Our results show that, when we consider single-message multiple multicasts over a quantum network
constructed with butterfly network blocks,
there is an important discrepancy between quantum and classical communication rates.
We have demonstrated that this discrepancy can be monotonically increased by adding additional blocks in parallel, up to a maximum
value of one bit/qubit per use and receiver for networks constructed using
identical erasure channels.

By exploiting inter-node classical communication in erasure
networks, we have shown that the discrepancy is increased more
rapidly. Additionally, in this case we observe a notable
discrepancy even when we add blocks in series and the number of
butterfly blocks separating senders from receivers is large.  By
adding further blocks in parallel to create a grid, we can
increase the discrepancy by more than 60\%, i.e., in the value of
the critical point at which the classical rate beats the quantum bound.

Our results demonstrate that duplicating certain existing
classical network structures containing butterfly blocks in order
to build quantum counterparts can result in significantly lower
performance. It may be possible to exploit this performance
discrepancy in order to create a system in which a quantum
communication cannot beat a classical equivalent. In this sense,
our results provide a theoretical guide with which to engineer
such a system.

\bigskip

\textbf{Acknowledgments}. This work has been supported by the
EPSRC via the ``UK Quantum Communications HUB'' (EP/M013472/1) and
by the European Union via the project ``Continuous Variable
Quantum Communications'' (CiViQ, no 820466).

\end{document}